\documentclass[epj]{webofc}
\usepackage[utf8]{inputenc}
\usepackage[varg]{txfonts}   % Web of Conferences font
\usepackage{booktabs,slashed}
\usepackage{xcolor}
\definecolor{darkred}{rgb}{0.4,0.0,0.0}
\definecolor{darkgreen}{rgb}{0.0,0.4,0.0}
\definecolor{darkblue}{rgb}{0.0,0.0,0.4}
\usepackage[bookmarks,linktocpage,colorlinks,
    linkcolor = darkred,
    urlcolor  = darkblue,
    citecolor = darkgreen]{hyperref}
\usepackage{subfigure}
\wocname{EPJ Web of Conferences}
\woctitle{Lattice2017}
\newcommand{\gino}{\lambda}
\newcommand{\bgino}{\bar{\lambda}}

\DeclareMathOperator{\Tr}{Tr}

\newcommand{\aetap}{\ensuremath{\text{a--}\eta'}}

\newcommand{\afn}{\ensuremath{\text{a--}f_0}}

\begin{document}
%%%%%%%%%%%%%%%%%%%%%%%%%%%%%%%%%%%%%%%%%%%%%%%%%%%%%%%%%%%%%%%%%%%%%%%%%%%%%
%
\selectlanguage{english}
%----------------------------------------------------------------------------
\title{%
Improved results for the mass spectrum of \protect{$\mathcal{N}=1$} supersymmetric SU(3) Yang-Mills theory 
}
%----------------------------------------------------------------------------
\author{%
\firstname{Sajid} \lastname{Ali}\inst{2} \and
\firstname{Georg} \lastname{Bergner}\inst{1,2}\fnsep\thanks{Speaker, \email{georg.bergner@uni-jena.de}} \and
\firstname{Henning} \lastname{Gerber}\inst{2} \and
\firstname{Pietro} \lastname{Giudice}\inst{2} \and
\firstname{Simon} \lastname{Kuberski}\inst{2} \and
\firstname{Gernot} \lastname{M\"unster}\inst{2} \and
\firstname{István} \lastname{Montvay}\inst{3} \and
\firstname{Stefano} \lastname{Piemonte}\inst{4} \and
\firstname{Philipp} \lastname{Scior}\inst{2} 
}
%----------------------------------------------------------------------------
\institute{%
Institute of Theoretical Physics, Friedrich-Schiller-University Jena, Max-Wien-Platz 1, D-07743 Jena, Germany
\and
University of M\"unster, Institute for Theoretical Physics, Wilhelm-Klemm-Str.~9, D-48149 M\"unster, Germany
\and
Deutsches Elektronen-Synchrotron DESY, Notkestr. 85, D-22603 Hamburg, Germany
\and
University of Regensburg, Institute for Theoretical Physics, Universit\"atsstr.~31, D-93040 Regensburg, Germany
}
%----------------------------------------------------------------------------
\abstract{%
This  talk summarizes the results of the DESY-Münster collaboration for $\mathcal{N}=1$ supersymmetric Yang-Mills theory with the gauge group SU(3). It is an updated status report with respect to our preliminary data presented at the last conference.  In order to control the lattice artefacts we have now considered a clover improved fermion action and different values of the gauge coupling.
}
%----------------------------------------------------------------------------
\maketitle
%----------------------------------------------------------------------------
\section{Supersymmetry on the lattice}
\label{intro}
Supersymmetric gauge theories are the main building blocks for supersymmetric extensions of the Standard Model. 
In addition they provide interesting new insights in the fundamental mechanisms of strongly interacting field theories. 
Non-perturbative numerical simulations on the lattice would mean an important progress in this field for two main reasons. 
They would provide a tool to investigate possible non-perturbative breaking mechanisms of supersymmetry.  A breaking mechanism for supersymmetry is an essential ingredient for supersymmetric
extensions of the Standard Model since the supersymmetric partner particles are not present in the observed particle spectrum. 

The second application of non-peturbative numerical tools in supersymmetric theories is related to theoretical considerations of supersymmetric theories. In some cases the 
extended symmetry allows for a better analytical understanding of supersymmetric theories. The calculation of the gluino condensate in $\mathcal{N}=1$ supersymmetric Yang-Mills theory and the 
gauge gravity dualities derived for theories with extended supersymmetry are an example. The final motivation for these investigations is to extend the new analytical insights and to relate them 
to more realistic theories. The results of non-perturbative numerical investigations provide checks for the reliability of the results and the extensions to a more general picture.

The main focus of our investigations is $\mathcal{N}=1$ supersymmetric Yang-Mills theory. In our first investigations we have mainly considered
the gauge group SU(2). We have measured the particle spectrum of the theory and observed the expected degeneracy of the supermultiplet of lowest states in this theory \cite{Bergner:2015adz}. Detailed considerations of 
chiral and continuum extrapolations as well as studies of the finite size effects were necessary to obtain this result. We have extended our studies now to the multiplet of excited states and the first
results are reported in a separate contribution to this conference \cite{Gerber}.

Our most recent investigations are numerical simulations of  $\mathcal{N}=1$ supersymmetric Yang-Mills theory with the gauge group SU(3). This theory is in some respects more realistic since it would correspond to 
the gauge part of supersymmetric QCD. We have reported the first results obtained with an unimproved Wilson fermion action at the last conference \cite{Ali:2016zke}. Simulations of the adjoint 
representation of SU(3) require much larger resources compared to the fundamental one. Therefore we are bound to rather small lattice sizes and we are not able to carry out the complete extrapolations that we 
have done for SU(2). We have found that with these kind of constraints the clover improved fermion action is a better choice than the Wilson one. In this talk we present our first results for the mass 
spectrum with this fermion action. They already indicate the expected formation of the supermultiplets of the lowest lying states as in the case of gauge group SU(2).
%----------------------------------------------------------------------------
\section{Supersymmetric Yang-Mills theory}\label{sec-1}
Supersymmetric Yang-Mills theory contains gluons and their supersymmetric partners, the gluinos. Supersymmetry requires the 
same number of degrees of freedom and the same representations of the fields. Therefore the gluinos ($\gino$) are Majorana fermions in the adjoint representation.
Taking into account an additional gluino mass term $m_g$ that breaks supersymmetry softly, the Lagrangean is
\begin{equation}
 \mathcal{L}=\Tr\left[-\frac{1}{2}
F_{\mu\nu}F^{\mu\nu}+i\bar{\gino}\slashed{D}\gino
-m_g\bgino\gino \right]\; .
\end{equation}
The theory is the counterpart of pure Yang-Mills theory with the usual properties like confinement of fundamental charges. 
On the other hand, it is in several respects similar to QCD since it contains fermionic fields. The primary observables are bound state masses. The states are represented by
flavor singlet mesonic operators, the scalar $\bgino\gino$ (\afn) and the pseudoscalar $\bgino\gamma_5\gino$ (\aetap), glueball operators, and gluino-glue operators.
The gluino-glue particles are a special feature of theories with fermions in the adjoint representation. Their interpolating operator is
\begin{align}
  \sum_{\mu,\nu} \sigma_{\mu\nu} \Tr\left[F^{\mu\nu} \lambda \right]\; .
\end{align}
These particles are the fermionic partners of the bosonic constituents of the supermultiplet of bound states.

Each massive chiral supermultiplet consists of a bosonic scalar, a bosonic pseudoscalar, and a fermionic particle with the same mass. There have been different predictions concerning
the low energy effective theory. A multiplet of mesons and the gluino-glue has been proposed in \cite{Veneziano:1982ah}, later on the lowest multiplet was conjectured
to be composed out of glueballs and the gluino-glue \cite{Farrar:1997fn}. It is an interesting question whether the expected formation of bound state supermultiplets can be 
confirmed by numerical lattice simulations. Further questions are related to the nature of the particles of the lowest and heavier multiplets and the 
splitting of the multiplets when a finite gluino mass is introduced.

There are several interesting properties of supersymmetric Yang-Mills theory. The Lagrangean is invariant under supersymmetry and chiral symmetry transformations. 
The chiral symmetry is not completely broken by the anomaly, but instead there is a remnant $Z_{2N_c}$ symmetry for the gauge group $\text{SU}(N_c)$. This remnant symmetry 
is spontaneously broken to $Z_2$ by the gluino condensate.

Simulations of $\mathcal{N}=1$ supersymmetric Yang-Mills theory are challenging for two basic reasons: the first one is the conceptual problem of any supersymmetric theory on the lattice. Supersymmetry is generically 
broken by the lattice discretization and can only be established in the continuum limit. The second one is a technical problem since the measurement of glueballs and 
singlet meson states usually provides a rather noisy signal and, consequently, rather large statistics is required at a rather fine lattice spacing.
%----------------------------------------------------------------------------
\section{Supersymmetric Yang-Mills theory on the lattice}\label{sec-2}
In our simulations we are using a Wilson plaquette action and a clover improved Wilson fermion action
\begin{align}
\mathcal{S}_L=\beta \sum_P\left(1-\frac{1}{N_c}\Re U_P\right) +\frac{1}{2}\sum_{xy} \bgino(x)
D_w(x,y)\gino(y)\; ,
\end{align}
where the Wilson-Dirac operator is 
\begin{align}
D_w(x,y)\gino(y)=\gino(x)-\kappa\sum_{\mu=1}^{4}\left[(1-\gamma_\mu)V_\mu(x)\gino(x)+(1+\gamma_\mu)V^\dag_\mu(x-\mu) \gino(x-\mu)\right]-\frac{c_{sw}}{4}\sigma_{\mu\nu}F^{\mu\nu}\gino(x)\; ,
\end{align}
with links in adjoint representation of SU(2) and SU(3): $(V_\mu)_{ab}=2\Tr[U_\mu^\dag T^a U_\mu T^b]$. The hopping parameter $\kappa$ is related to the bare gluino mass $\kappa=\frac{1}{2m+8}$. The field strength $F_{\mu\nu}$ is represented by clover plaquette operators. We have calculated the tuning of the clover coefficient $c_{sw}$ in lattice 
perturbation theory \cite{Musberg:2013foa}. In the current investigations we are using the one-loop clover coefficient.

The Wilson fermion action breaks chiral symmetry and supersymmetry. Both symmetries can be restored in the continuum limit via the fine-tuning of a single parameter, the bare gluino mass \cite{Curci:1986sm}. This Veneziano-Curci scenario is a theoretical prediction. Its applicability has to be verified in numerical simulations. 

The fine-tuning of the gluino mass has to be done with respect to a signal for the broken chiral symmetry or supersymmetry. The chiral symmetry is subject to the anomaly and hence the signal for its restoration is in principle rather difficult to identify. However, in the framework of partially quenched chiral perturbation theory, the signal for the chiral symmetry restoration can be related to the adjoint pion mass \cite{Munster:2014cja}. This mass does not belong to a physical particle of the theory.
The second possibility is a tuning with respect to the supersymmetric Ward identities, which have, however, a less precise signal than the adjoint pion mass. We are therefore taking both tunings into account: the extrapolations towards the chiral limit are based on the adjoint pion mass, but these are checked against the signal for the supersymmetric Ward identities. The details of these tunings are presented in a separate contribution \cite{Ali}. Note that an interesting alternative approach is the consideration of Domain-Wall or overlap fermions, which don't require a tuning for the restoration of chiral symmetry \cite{Kaplan:1999jn}.

We have confirmed in large scale simulations of SU(2) supersymmetric Yang-Mills theory that the signals for a restored supersymmetry, in particular in terms of the formation of supermultiplets, are observed in the continuum limit. The lattice artefacts are a considerable obstacle in these investigations, and we have also done a detailed check of finite size effects. We have obtained the masses of the lowest supermultiplet in the continuum limit \cite{Bergner:2015adz}. In further studies we have investigated the finite temperature properties of supersymmetric Yang-Mills theory. We have found an interesting coincidence between the chiral and the deconfinement transition \cite{Bergner:2014saa}. We have also considered the case where the boundary conditions for the fermions are changed from the thermal to the periodic case \cite{Bergner:2014dua}. In this case there is a disappearance of the deconfinement transition, in agreement with the theoretical predictions.

Recently, we have extended our studies in several ways. We have considered new operators and made the first preliminary study of the multiplet of the first excited states, which is presented in a separate contribution \cite{Gerber}. Moreover, we have started to investigate the gauge group SU(3) instead of the simpler SU(2) case. First results have been presented at the last lattice conference \cite{Ali:2016zke}. The adjoint representation of SU(3) requires a considerably larger amount of resources than the fundamental one or the SU(2) case. Therefore we are not able to scan the same range of lattice spacings and volumes as in our previous  investigations of SU(2) supersymmetric Yang-Mills theory. In the earlier investigations we have observed that the clover improvement leads to a significant reduction of the lattice artefacts in terms of the splitting between the masses of the lightest supermultiplet \cite{Bergner:2015adz}. Therefore we have considered a clover improved action in our most recent SU(3) supersymmetric Yang-Mills theory simulations.

%----------------------------------------------------------------------------
\section{Results for $\mathcal{N}=1$ supersymmetric SU(3) Yang-Mills theory}\label{sec-3}
\begin{figure}
\centerline{\includegraphics[width=0.7\textwidth]{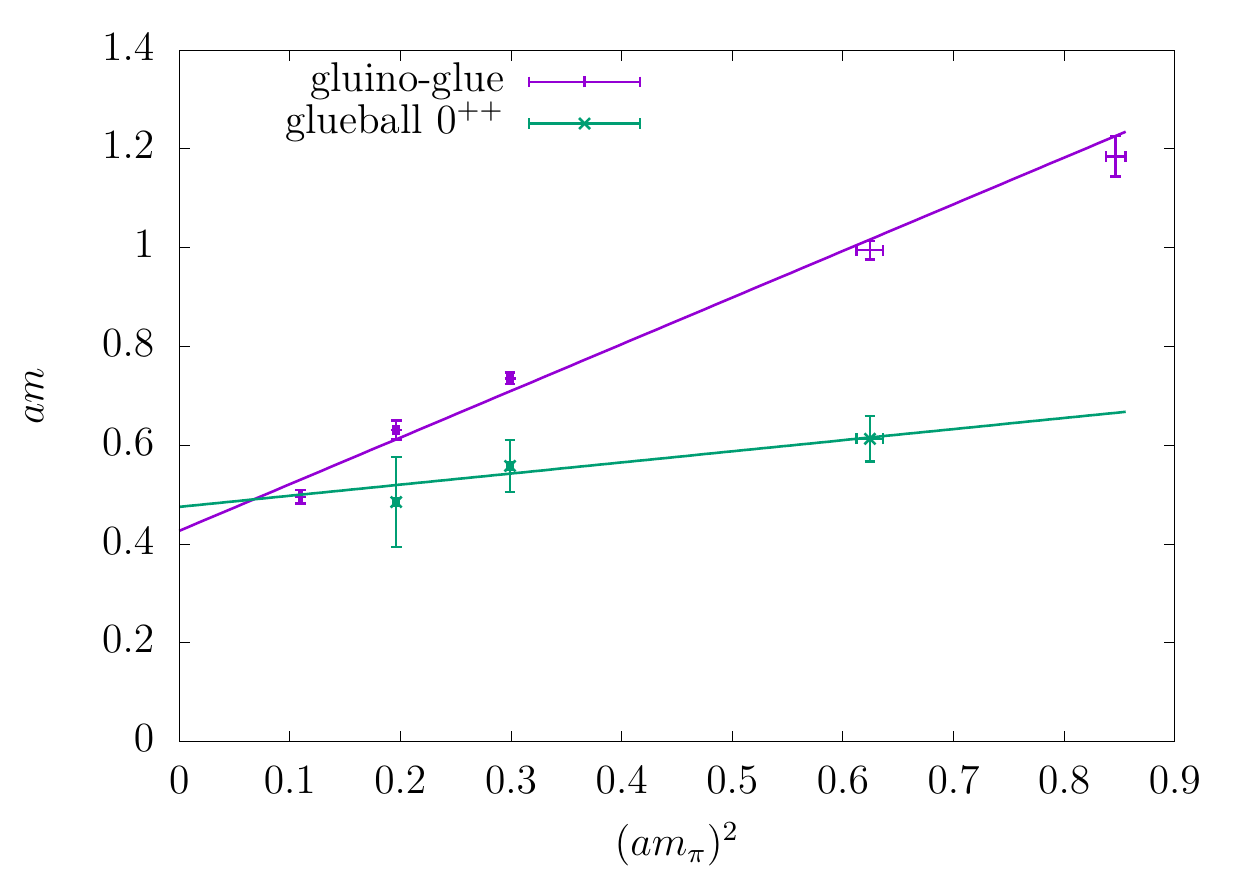}}
\caption{The particle spectrum of $\mathcal{N}=1$ supersymmetric SU(3) Yang-Mills theory. This figure shows the masses of the fermonic gluino-glue particle and the $0^{++}$ glueball as a function of the square of the adjoint pion mass.  The chiral limit, corresponding to zero adjoint pion mass, is linearly extrapolated. All masses are presented in lattice units.}
\label{fig:gggb}
\end{figure}
\begin{figure}
\centerline{\includegraphics[width=0.7\textwidth]{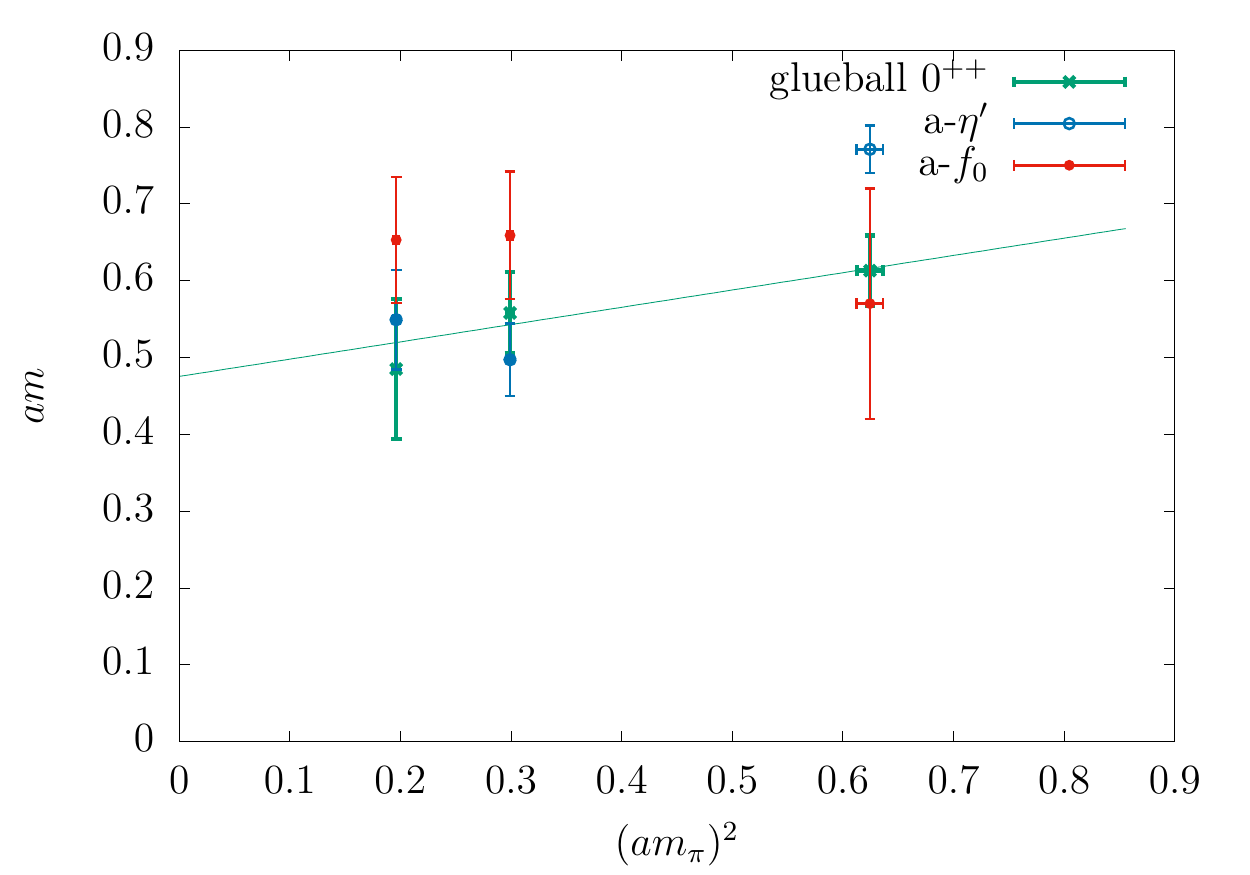}}
\caption{The particle spectrum of $\mathcal{N}=1$ supersymmetric SU(3) Yang-Mills theory. This figure shows our preliminary results for the masses of the mesonic particles mesonic particles in comparison to the $0^{++}$ glueball mass. For comparison the linear extrapolation of the glueball mass towards the chiral limit has been added to the plot.}
\label{fig:mesgb}
\end{figure}
We have investigated several different lattice spacings with the clover improved lattice action, but currently we have only obtained reasonable statistics with our ensembles at a rather coarse lattice spacing, corresponding to $a\approx 0.08$~fm in QCD units. The scale setting has been determined from the Sommer scale $r_0$ of the static quark-antiquark potential and the $w_0$ scale of the Gradient flow. In our previous simulations with the gauge group SU(2) we have observed a quite large (around $50\%$) mass gap between the bosonic and fermionic constituents of the lowest supermultiplet when a plain Wilson fermion action is considered at this rather coarse lattice spacing. In the clover improved case at the same lattice spacing, the particle masses have been already degenerate within errors. We expect a similar situation for the gauge group SU(3). 

We have confirmed that the square of the adjoint pion mass is, to a reasonable approximation, consistent with the remnant gluino mass determined from the supersymmetric Ward identities \cite{Ali}. We have determined the gluino-glue and glueball masses. In addition we have measured the singlet mesons according to the methods presented in \cite{Bergner:2011zz}. All these methods have already been tested and optimized in our previous analysis of the gauge group SU(2).

The masses of the bosonic $0^{++}$ glueball and the fermionic gluino-glue are shown in Figure~\ref{fig:gggb} as a function of the square of the adjoint pion mass. These preliminary results indicate a restoration of the degeneracy between the bosonic and fermionic masses in the chiral limit. 

The determined mesonic masses are so far not of the same accuracy as those of the glueball and the gluino-glue, see Figure \ref{fig:mesgb}. The scalar \afn\ meson has the same quantum numbers as the $0^{++}$ glueball and a mixing between these states is expected. This is in accordance with the fact that in our results both masses agree within errors. A similar mixing is expected for the pseudoscalar \aetap\ meson and the $0^{-+}$ glueball, but so far we did not get a sufficient signal for the mass of the pseudoscalar glueball. Our results for SU(2) supersymmetric Yang-Mills theory have shown that the mesonic \aetap\ operator has a much better overlap with the ground state than the glueball. Consequently, we represent the pseudoscalar constituent of the lightest supermultiplet only by the mesonic \aetap. Our results for SU(3) $\mathcal{N}=1$ supersymmetric SU(3) Yang-Mills theory show hence for the first time the degeneracy of the particle masses of the lightest supermultiplet consisting of a scalar, a pseudoscalar, and a fermionic component.

%----------------------------------------------------------------------------
\section{Conclusions}\label{sec-4}
We have presented the results of our lattice simulations of $\mathcal{N}=1$ supersymmetric SU(3) Yang-Mills theory. Our results show a first indication of degeneracy in the bound state particle spectrum consistent with the formation of a supermultiplet of a scalar, a pseudoscalar, and a fermionic particle. This generalizes our earlier investigations for the gauge group SU(2).

The effect of the clover improvement and lattice artefacts are similar to the SU(2) case: a rather fine lattice is required to obtain the degeneracy of the bound state masses when a plain Wilson-Dirac operator is used. Therefore we have not observed the expected multiplet formation in our earlier results \cite{Ali:2016zke}. The clover improved fermion action leads to a considerable reduction of lattice artefacts and the signals for restored supersymmetry are consequently obtained already on a rather coarse lattice.  

%----------------------------------------------------------------------------
\section*{Acknowledgements}

The authors gratefully acknowledge the Gauss Centre for Supercomputing (GCS) for providing computing time for a GCS Large-Scale Project on the GCS share of the supercomputer JUQUEEN and JURECA at J\"ulich Supercomputing Centre (JSC) and on the supercomputer SuperMUC at Leibniz Computing Centre (LRZ). GCS is the alliance of the three national supercomputing centres HLRS (Universität Stuttgart), JSC (Forschungszentrum J\"ulich), and LRZ (Bayerische Akademie der Wissenschaften), funded by the German Federal Ministry of Education and Research (BMBF) and the German State Ministries for Research of Baden-Württemberg (MWK), Bayern (StMWFK) and Nordrhein-Westfalen (MIWF). Further computing time has been provided on the compute cluster PALMA of the University of M\"unster.

%%%%%%%%%%%%%%%%%%%%%%%%%%%%%%%%%%%%%%%%%%%%%%%%%%%%%%%%%%%%%%%%%%%%%%%%%%%%%
\end{document}